\documentclass[english,prd,twocolumn,superscriptaddress,nofootinbib,preprintnumbers]{revtex4}
\usepackage[latin1]{inputenc}
\usepackage{graphicx}
\usepackage{amssymb}
\usepackage{epstopdf}
\usepackage{amsmath}
\usepackage{amsmath}
\usepackage{bm}
\makeatletter

\usepackage{amsfonts}
\usepackage{dcolumn}
\usepackage{hyperref}

\def\be{\begin{equation}}
\def\ee{\end{equation}}
\def\ba{\begin{eqnarray}}
\def\ea{\end{eqnarray}}
\def\bs{\begin{subequations}}
\def\es{\end{subequations}}

\newcommand{\rd}{{\rm d}}

\def \om   {\Omega_{\rm 0m}}

\def \ob  {\Omega_b^{(0)}}

\def \omms  {\Omega_m^{(0)}}

\def \omk   {\Omega_K^{(0)}}
\def \omr   {\Omega_r^{(0)}}

\def \om  {\Omega_m^{(0)}}

\newcommand{\newc}{\newcommand}
\newc{\lcdm }{$\Lambda$CDM }

\newc{\ra}{\rightarrow}
\newc{\lra}{\leftrightarrow}
\newc{\lsim}{\buildrel{<}\over{\sim}}
\newc{\gsim}{\buildrel{>}\over{\sim}}
\usepackage{color}

\makeatother

\usepackage{babel}
\makeatother
\begin{document}

\title{Observational constraints on Galileon cosmology}

\author{Savvas Nesseris}
\affiliation{The Niels Bohr International Academy and DISCOVERY Center, The Niels Bohr
Institute, Blegdamsvej 17, DK-2100, Copenhagen \O, Denmark}
\email{nesseris@nbi.dk}

\author{Antonio De Felice}
\affiliation{Department of Physics, Faculty of Science, Tokyo University of Science,
1-3, Kagurazaka, Shinjuku-ku, Tokyo 162-8601, Japan}
\email{defelice@rs.kagu.tus.ac.jp}

\author{Shinji Tsujikawa}
\affiliation{Department of Physics, Faculty of Science, Tokyo University of Science,
1-3, Kagurazaka, Shinjuku-ku, Tokyo 162-8601, Japan}
\email{shinji@rs.kagu.tus.ac.jp}

\begin{abstract}

We study the cosmology of a covariant Galileon field $\phi$ with five covariant
Lagrangians and confront this theory with the most recent cosmological probes:
the type Ia supernovae data (Constitution and Union2 sets), cosmic microwave
background (WMAP7) and the baryon acoustic oscillations  (SDSS7).
In the Galileon cosmology with a late-time de Sitter attractor,
there is a tracker that attracts solutions with different initial conditions
to a common trajectory.
Including the cosmic curvature $K$, we place observational constraints
on two distinct cases: (i) the tracker, and (ii) the generic solutions
to the equations of motion. We find that the tracker solution can be
consistent with the individual observational data,
but it is disfavored by the combined data analysis.
The generic solutions fare quite well when a non-zero curvature parameter
$\omk$ is taken into account, but the Akaike and Bayesian information criteria
show that they are not particularly favored over
the $\Lambda$CDM model.

\end{abstract}

\date{\today}

\maketitle

\section{Introduction}

The quest to distinguish between a cosmological constant and
dynamical dark energy models has been one of the main topics
in cosmology. The equation of state $w_{\rm DE}$ of dark energy
can be constrained not only by the Supernovae Ia (SN Ia)
data \cite{SNearly} but by the observations of Cosmic
Microwave Background (CMB) \cite{CMBearly}
and Baryon Acoustic Oscillations (BAO) \cite{BAO1,Percival:2009xn}.
For constant $w_{\rm DE}$ the combined data analysis of
CMB$+$BAO$+$SN Ia by the WMAP group has given a tight constraint
$w_{\rm DE}=-0.980 \pm 0.053$ (68 \% confidence level) in the
flat Universe \cite{Komatsu:2010fb}. However, the present observations
still allow a large variation of the dark energy equation of state in terms
of the redshift $z$.
Moreover, inclusion of the cosmic curvature further weakens
the constraints on $w_{\rm DE}(z)$.

Over the past decade, many dynamical dark energy models have
been proposed as an alternative to the cosmological
constant (see Refs.~\cite{review} for review).
They are broadly classified into two
classes--(i) Modified matter
models, and (ii) Modified gravity models.
In the class (i) the accelerated expansion of the Universe
is induced by a modified matter source,
whereas in the class (ii) the large-distance
modification of gravity gives rise to the cosmic acceleration.
The representative model of the class (i) is quintessence based
on a minimally coupled scalar field \cite{quin}, but in general
there is a degeneracy around $w_{\rm DE}=-1$ if
we constrain the quintessence potential from observations.

The modified gravity models proposed so far consist of
$f(R)$ gravity \cite{fR}, scalar-tensor theory \cite{stensor},
the Dvali-Gabadadze-Porrati (DGP) braneworld \cite{DGP}
model, the Gauss-Bonnet gravity \cite{fG}, $f(R,{\cal G})$
gravity \cite{fRG}, and so on.
In general we need to recover the General Relativistic behavior
in the region of high density for the consistency
with solar-system experiments \cite{Will},
while the large-distance modification of gravity leads to
the cosmic acceleration today.
Moreover we require that the models are free from
ghosts and instabilities \cite{DeFelice}.
For example, the DGP model is plagued by
the ghost problem \cite{DGPghost} in addition to the
incompatibility with observational constraints \cite{DGPobser}.
The dark energy models in which the
Lagrangian includes a general function $f$ of the Gauss-Bonnet
term ${\cal G}$ also result in violent instabilities for
small-scale perturbations \cite{fGins}.
In $f(R)$ gravity and scalar-tensor
theory, the functions $f(R)$ or the field potentials need to be
carefully designed to satisfy the above-mentioned
demands \cite{fRviable}.

In the DGP model a brane-bending mode $\phi$
(i.e. longitudinal graviton)
gives rise to a field self-interaction of the form
$\square \phi (\partial^{\mu} \phi \partial_{\mu} \phi)$
through the mixing with the transverse graviton \cite{DGPnon}.
This allows the decoupling of the field $\phi$ from
gravitational dynamics in the local region by the so-called
Vainshtein mechanism \cite{Vainshtein}.
Then the General Relativistic behavior can be recovered
within a radius larger than the solar-system scale.
The self-interaction $\square \phi (\partial^{\mu} \phi \partial_{\mu} \phi)$ satisfies the Galilean symmetry
$\partial_\mu\phi\to\partial_\mu\phi+b_\mu$
in the flat space-time.
While the DGP model suffers from the ghost problem,
the extension of the field self-interaction to more
general forms satisfying the Galilean symmetry
may allow us to avoid the appearance of ghosts.

Nicolis {\it et al.} \cite{Nicolis} showed that there are
only five field Lagrangians ${\cal L}_i$ ($i=1,\cdots, 5$)
that respect the Galilean symmetry
in the Minkowski background.
These terms lead to only the second-order field equations
and hence we do not need to worry about extra degrees of
freedom coming from higher-order derivatives.

If the Lagrangians ${\cal L}_i$ are varied in the curved
space-time, the terms ${\cal L}_{4,5}$ give
the field equations higher than the second-order.
Deffayet {\it et al.} \cite{Deffayet} derived the covariant Lagrangians
${\cal L}_i$ ($i=1,\cdots, 5$) that result in only the second-order
equations, while recovering the Galilean symmetry in the
Minkowski space-time. This can be achieved by introducing
field-derivative couplings with the Ricci scalar $R$ and
the Einstein tensor $G_{\nu \rho}$ in the expression of
${\cal L}_{4,5}$. Since the existence of those terms
affects the effective gravitational coupling, the Galileon
gravity based on the covariant Lagrangians
${\cal L}_i$ ($i=1,\cdots, 5$) can be classified as one of
modified gravitational theories.

The full cosmological dynamics
including the terms up to ${\cal L}_{5}$ have been
studied by two of the present authors
in the flat Friedmann-Lema\^{i}tre-Robertson-Walker
(FLRW) background \cite{DT2,DT3}
(see also Refs.~\cite{JustinGal}-\cite{Mota10} for related works).
While the field is nearly frozen during the early epoch
through the cosmological Vainshtein mechanism,
it begins to evolve at late times to lead to the acceleration
of the Universe. The stable dS solution can be realized by
a constant field velocity.

Refs.~\cite{DT2,DT3} have shown that, for the covariant
Galileon theory having dS attractors, cosmological solutions
with different initial conditions converge to a common
trajectory-- a tracker solution.
Moreover the background cosmological dynamics along the
tracker can be known analytically in terms of the redshift $z$.
The dark energy equation of state exhibits the peculiar phantom-like
evolution: $w_{\rm DE}=-7/3$ (radiation era),
$w_{\rm DE}=-2$ (matter era), and $w_{\rm DE}=-1$ (dS era).
Note that this does not imply the appearance of ghosts.
In fact the viable model parameter space has been found
in Refs.~\cite{DT2,DT3} from the conditions to avoid ghosts and
Laplacian instabilities of scalar and tensor perturbations.

In this paper we place observational constraints on the covariant
Galileon gravity using the observational data of SN Ia,
the CMB shift parameters, and BAO.
In particular we derive a convenient analytic formula for
the tracker evolution by including the cosmic curvature $K$
and test the viability of such a solution.
In general the cosmological dynamics start from the regime away from
the tracker, depending on the model parameters and initial conditions.
We shall also study such general cases and search for the
model parameter space consistent with observational constraints.

\section{Galileon cosmology}

The covariant Galileon gravity is described
by the action \cite{Deffayet}
\begin{equation}
S=\int {\rm d}^4 x \sqrt{-g}\,\left[ \frac{M_{\rm pl}^2}{2}R+
\frac12 \sum_{i=1}^5 c_i {\cal L}_i \right]
+\int {\rm d}^4 x\, {\cal L}_{M}\,,
\label{action}
\end{equation}
where $g$ is a determinant of the metric tensor $g_{\mu \nu}$,
$M_{\rm pl}$ is the reduced Planck mass, and
$c_i$ are constants.
The covariant Lagrangians ${\cal L}_i$ ($i=1, \cdots, 5$)
that respect the Galilean symmetry in the limit
of the Minkowski space-time are given by
\begin{eqnarray}
& & {\cal L}_1=M^3 \phi\,,\quad
{\cal L}_2=(\nabla \phi)^2\,,\quad
{\cal L}_3=(\square \phi) (\nabla \phi)^2/M^3\,, \nonumber \\
& & {\cal L}_4=(\nabla \phi)^2 \left[2 (\square \phi)^2
-2 \phi_{;\mu \nu} \phi^{;\mu \nu}-R(\nabla \phi)^2/2 \right]/M^6,
\nonumber \\
& & {\cal L}_5=(\nabla \phi)^2 [ (\square \phi)^3
-3(\square \phi)\,\phi_{; \mu \nu} \phi^{;\mu \nu} \nonumber \\
& &~~~~~~~+2{\phi_{;\mu}}^{\nu} {\phi_{;\nu}}^{\rho}
{\phi_{;\rho}}^{\mu}
-6 \phi_{;\mu} \phi^{;\mu \nu}\phi^{;\rho}G_{\nu \rho} ]
/M^9\,,
\label{lag}
\end{eqnarray}
where a semicolon represents a covariant derivative,
$M$ is a constant having a dimension of mass, and $G_{\nu \rho}$ is the Einstein tensor.
For the matter Lagrangian ${\cal L}_{M}$
we take into account perfect fluids of
radiation (energy density $\rho_r$, equation of state $w_r=1/3$) and non-relativistic matter
(energy density $\rho_m$, equation of
state $w_m=0$).

We consider the FLRW space-time with the line element
\begin{equation}
{\rm d}s^2=-{\rm d}t^2+a^2(t) \left[
\frac{{\rm d} r^2}{1-Kr^2}+r^2 ({\rm d}\theta^2+
\sin^2 \theta\,{\rm d}\phi^2) \right]\,,
\label{metric}
\end{equation}
where $a(t)$ is the scale factor with the cosmic time $t$.
The closed, flat, and open geometries correspond to
$K>0$, $K=0$, and $K<0$, respectively.
Variation of the action (\ref{action}) with respect to $g_{\mu \nu}$
leads to the following equations of motion
\begin{eqnarray}
& & 3M_{\rm pl}^2 H^2=\rho_{\rm DE}+\rho_m+\rho_r+\rho_K\,,
\label{basic1} \\
& & 3M_{\rm pl}^2 H^2+2M_{\rm pl}^2 \dot{H}=-P_{\rm DE}
-\rho_r/3+\rho_K/3\,,
\label{basic2}\\
& & \dot{\rho}_m+3H\rho_m=0\,,
\label{basic3}\\
& & \dot{\rho}_r+4H\rho_r=0\,,
\label{basic4}
\end{eqnarray}
where $\rho_K \equiv -3KM_{\rm pl}^2/a^2$,
a dot represents a derivative with respect to $t$,
$H=\dot{a}/a$ is the Hubble parameter, and
\begin{eqnarray}
\rho_{\rm DE} &\equiv& -c_1 M^3 \phi/2-c_2 \dot{\phi}^2/2
+3c_3 H \dot{\phi}^3/M^3 \nonumber \\
& &-45 c_4 H^2 \dot{\phi}^4/(2M^6)
+21c_5 H^3 \dot{\phi}^5/M^9,\\
P_{\rm DE} &\equiv&  c_1 M^3 \phi/2-c_2 \dot{\phi}^2/2
-c_3 \dot{\phi}^2 \ddot{\phi}/M^3 \nonumber \\
& &+3c_4 \dot{\phi}^3 [8H\ddot{\phi} +(3H^2+2\dot{H})
\dot{\phi}]/(2 M^6) \nonumber \\
& & -3c_5 H \dot{\phi}^4 [5H \ddot{\phi}+2(H^2+\dot{H})
\dot{\phi} ]/M^9\,.
\end{eqnarray}

Note that Eqs.~(\ref{basic1}) and (\ref{basic2}) are the generalization of those derived in Ref.~\cite{DT2} with an account of the cosmic curvature $K$.
The closed-form equations for $\ddot{\phi}$ and $\dot{H}$
can be derived by taking a time-derivative of
Eq.~(\ref{basic1}) and by combining it with Eq.~(\ref{basic2}).
Since we are interested in the case where the late-time cosmic
acceleration is realized by the field kinetic energy, we
set $c_1=0$ in the following discussion.
In this case the only solution in the Minkowski background ($H=0$)
corresponds to $\dot{\phi}=0$ for $c_2 \neq 0$.

The dS solution ($H=H_{\rm dS}={\rm constant}$) can
be present for $\dot{\phi}=\dot{\phi}_{\rm dS}={\rm constant}$.
We normalize the mass $M$ to be $M^3=M_{\rm pl}H_{\rm dS}^2$,
which gives $M \approx 10^{-40}M_{\rm pl}$ for
$H_{\rm dS} \approx 10^{-60}M_{\rm pl}$.
Defining $x_{\rm dS} \equiv \dot{\phi}_{\rm dS}/(H_{\rm dS}M_{\rm pl})$,
Eqs.~(\ref{basic1}) and (\ref{basic2}) give the following relations
at the dS point:
\begin{eqnarray}
& & c_2 x_{\rm dS}^2=6+9\alpha-12\beta\,,
\label{ds1} \\
& & c_3 x_{\rm dS}^3=2+9\alpha-9\beta\,,
\label{ds2}
\end{eqnarray}
where
\begin{equation}
\alpha \equiv c_4 x_{\rm dS}^4\,,\qquad
\beta \equiv c_5 x_{\rm dS}^5\,.
\end{equation}
The use of $\alpha$ and $\beta$ is convenient because
the coefficients of physical quantities and dynamical equations
can be expressed by those variables.
We note that the relations (\ref{ds1}) and (\ref{ds2}) are not
subject to change under the rescaling $x_{\rm dS} \to \gamma x_{\rm dS}$
and $c_i \to c_i/\gamma^i$, where $\gamma$ is a real constant.
Hence the rescaled choice of $c_i$ can provide the same physics.
If we use the parameters $\alpha$ and $\beta$, such apparent different cases
can be treated in a unified way.
In Refs.~\cite{DT2,DT3} the authors derived
the viable parameter space on the $(\alpha,\beta)$ plane in which
the conditions for the avoidance of ghosts and Laplacian instabilities of
scalar and tensor perturbations are satisfied for $K=0$.

In order to study the cosmological dynamics, it is useful to define
the following dimensionless variables:
\begin{equation}
r_1 \equiv \frac{\dot{\phi}_{\rm dS}H_{\rm dS}}{\dot{\phi}H}\,,
\qquad
r_2 \equiv \frac{1}{r_1} \left( \frac{\dot{\phi}}{\dot{\phi}_{\rm dS}}
\right)^4\,.
\label{r1r2def}
\end{equation}
At the dS solution one has $r_1=1$ and $r_2=1$.
We define the dark energy density parameter
\begin{eqnarray}
\Omega_{\rm DE} &\equiv& \frac{\rho_{\rm DE}}{3M_{\rm pl}^2 H^2}
\nonumber \\
&=& -(2+3\alpha-4\beta)r_1^3r_2/2+(2+9\alpha-9\beta)r_1^2r_2
\nonumber \\
&& -15\alpha r_1 r_2/2+7\beta r_2\,,
\label{OmeDE}
\end{eqnarray}
where we have used Eqs.~(\ref{ds1}) and (\ref{ds2}).
Then Eq.~(\ref{basic1}) can be written as
\begin{equation}
\Omega_{\rm DE}+\Omega_m+\Omega_r+\Omega_K=1\,,
\label{con}
\end{equation}
where $\Omega_m \equiv \rho_m/(3M_{\rm pl}^2 H^2)$,
$\Omega_r \equiv \rho_r/(3M_{\rm pl}^2 H^2)$, and
$\Omega_K \equiv \rho_K/(3M_{\rm pl}^2 H^2)=-K/(aH)^2$.

From Eqs.~(\ref{basic1})-(\ref{basic4}) we obtain the following
autonomous equations for the variables $r_1$, $r_2$,
$\Omega_r$, and $\Omega_K$:
\begin{widetext}
\begin{eqnarray}
\label{eq:DRr1}
\hspace{-0.5cm}
r_1' &=& \frac{1}{\Delta} \left(r_1-1\right)
r_1 \left[ r_1 \left(r_1 (-3 \alpha +4 \beta -2)
+6 \alpha -5 \beta \right)-5 \beta \right] \nonumber\\
&&{}\times \left[ 2 \left(\Omega _r-\Omega_K+9\right)
+3 r_2 \left( r_1^3 (-3 \alpha +4\beta -2)+
2 r_1^2 (9 \alpha -9 \beta +2)-15 r_1 \alpha
+14 \beta \right)\right]\,,\\
\label{eq:DRr2}
\hspace{-0.5cm}
r_2' &=& -\frac{1}{\Delta}
[ r_2 (6 r_1^2 (r_2 (45 \alpha ^2-4 (9 \alpha +2) \beta
+36 \beta^2)-(\Omega_r-\Omega_K-7) (9 \alpha -9 \beta +2))
+r_1^3 (-2(\Omega_r-\Omega_K+33) \nonumber \\ & &
\times (3 \alpha -4 \beta +2)
-3 r_2 (-2 (201 \alpha +89) \beta +15\alpha
(9 \alpha +2)+356 \beta ^2))
-3 r_1 \alpha (-28 \Omega _r+28\Omega_K+123 r_2
\beta +36) \nonumber \\
& &+10 \beta (-11 \Omega _r+11\Omega_K
+21 r_2 \beta -3) +3r_1^4 r_2
(9\alpha ^2-30 \alpha (4 \beta +1)+2 (2-9 \beta )^2)
+3r_1^6 r_2 (3 \alpha -4 \beta+2)^2 \nonumber \\
& &+3 r_1^5 r_2
(9 \alpha -9 \beta +2) (3 \alpha -4 \beta +2))], \\
\label{eq:DRr3}
\hspace{-0.5cm}
\Omega_r' &=& \Omega_r \left( -4 -2\frac{H'}{H} \right)\,,\\
\hspace{-0.5cm}
\label{eq:DRr4}
\Omega_K' &=& \Omega_K \left( -2 -2\frac{H'}{H} \right)\,,
\end{eqnarray}
where a prime represents a derivative with respect to
$N=\ln a$, and
\begin{eqnarray}
\Delta &\equiv & 2 r_1^4 r_2 [ 72 \alpha ^2+30 \alpha
(1-5 \beta )+(2-9 \beta )^2 ]+4 r_1^2
[ 9 r_2 (5 \alpha ^2+9 \alpha \beta +(2-9 \beta ) \beta )
+2 (9 \alpha -9 \beta +2) ] \nonumber\\
&&+4 r_1^3 [ -3 r_2 \left(-2 (15 \alpha +1) \beta
+3 \alpha  (9 \alpha +2)+4 \beta ^2\right)-3 \alpha
+4 \beta -2]-24 r_1 \alpha  (16 r_2 \beta +3)
+10\beta (21 r_2 \beta +8)\,.
\end{eqnarray}
\end{widetext}
The Hubble parameter follows from the equation
\begin{equation}
\frac{H'}{H}=-\frac{5r_1'}{4r_1}-\frac{r_2'}{4r_2}\,.\label{HN}
\end{equation}

The solutions to Eqs.~(\ref{eq:DRr3}) and (\ref{eq:DRr4})
are given by $\Omega_r (N)=\Omega_r^{(0)} e^{-4N}H_0^2/H^2(N)$
and $\Omega_K (N)=\Omega_K^{(0)} e^{-2N}H_0^2/H^2(N)$
respectively, where the subscript ``(0)'' represents the values
today ($N=0$ and $a=1$).
Hence these variables are related with each other via
\begin{equation}
\Omega_K(N)=\Omega_K^{(0)}
(\Omega_r(N)/\Omega_r^{(0)})e^{2N}\,.
\label{Omere}
\end{equation}
\subsection{Tracker solution}

Equation (\ref{eq:DRr1}) shows that there is an equilibrium
point characterized by
\begin{equation}
r_1=1\,,
\label{tracker}
\end{equation}
along which
\begin{equation}
\Omega_{\rm DE}=r_2\,,
\label{Omedetra}
\end{equation}
Originally the existence of the tracker solution (\ref{tracker})
was found in Refs.~\cite{DT2,DT3} for the flat Universe ($K=0$).
The above results show that the tracker is also present for $K \neq 0$.

The epoch at which the solutions reach the tracking regime
$r_1 \simeq 1$ depends on model parameters and initial conditions.
The approach to this regime occurs later for smaller initial values of $r_1$.
If $r_1 \lesssim 2$ initially, numerical simulations show
that the solutions approach the tracker with the
late-time cosmic acceleration. Meanwhile,
for the initial conditions with $r_1 \gtrsim 2$,
the dominant contribution to $\Omega_{\rm DE}$
comes from the Lagrangian ${\cal L}_2$, so that
the field energy density decreases rapidly as in
the standard massless scalar field.
In the latter case the solutions do not get out of
the matter era that starts from the
radiation-matter equality \cite{DT2,DT3}.

{}From Eqs.~(\ref{eq:DRr2})-(\ref{eq:DRr4})
the variables $r_2$, $\Omega_r$, and $\Omega_K$ satisfy
the following equations along the tracker:
\begin{eqnarray}
r_2' &=& \frac{2 r_2 \left( 3-3r_2+\Omega _r-\Omega_K \right)}
{1+r_2}\,,
\label{r2eq} \\
\Omega_r' &=& \frac{\Omega _r
\left(\Omega _r-\Omega_K-1-7 r_2\right)}{1+r_2}\,,
\label{Omereq} \\
\Omega_K' &=& \frac{\Omega_K
\left(\Omega _r-\Omega_K+1-5r_2\right)}{1+r_2}\,.
\label{Omekeq}
\end{eqnarray}
These equations do not have any dependence on $\alpha$ and $\beta$.
Combining Eqs.~(\ref{r2eq}) and (\ref{Omereq}), we obtain
\begin{equation}
\frac{r_2'}{r_2}=8+\frac{2\Omega_r'}{\Omega_r}\,.
\end{equation}
Integrating this equation, it follows that
\begin{equation}
r_2=d_1 a^8 \Omega_r^2\,,
\label{r2eq2}
\end{equation}
where $d_1$ is a constant.

{}From Eqs.~(\ref{Omereq}) and (\ref{Omekeq}) we have
\begin{equation}
\frac{\Omega_K'}{\Omega_K}-\frac{\Omega_r'}{\Omega_r}=2\,,
\end{equation}
which is integrated to give
\begin{equation}
\frac{\Omega_K}{\Omega_r}=d_2 a^2\,,\quad
{\rm with} \quad
d_2=\frac{\Omega_K^{(0)}}{\Omega_r^{(0)}}\,.
\label{Omera}
\end{equation}

Substituting Eqs.~(\ref{r2eq2}) and (\ref{Omera})
into Eq.~(\ref{Omereq}), we obtain the following
integrated solution
\begin{equation}
\label{eq:omsol}
\Omega_r=\frac{-1+d_3 a-d_2 a^2+\sqrt{4d_1 a^8+
(-1+d_3 a-d_2 a^2)^2}}{2d_1 a^8}\,,
\end{equation}
where $d_3$ is another constant.
Note that another solution of $\Omega_r$
(i.e. minus sign in front of the square root)
is not cosmologically viable, because of
the divergence of $\Omega_r$ as $a \to 0$.
The density parameter (\ref{eq:omsol}) evolves as
$\Omega_r \simeq 1+d_3a$ in the early time ($a \ll 1$).
This demands the condition $d_3<0$
provided that $\Omega_{\rm DE}>0$.

Using the density parameters $\Omega_m^{(0)}$,
$\Omega_r^{(0)}$,
and $\Omega_K^{(0)}$ today, the constants $d_1$ and
$d_3$ can be expressed as
\begin{equation}
d_1=\frac{1-\Omega_m^{(0)}-\Omega_r^{(0)}-\Omega_K^{(0)}}
{(\Omega_r^{(0)})^2}\,,\qquad
d_3=-\frac{\Omega_m^{(0)}}{\Omega_r^{(0)}}\,,
\label{d1d3}
\end{equation}
where we have used Eqs.~(\ref{con}), (\ref{Omedetra}), and (\ref{eq:omsol}).
In the high-redshift regime ($z \gg 1$) the density parameters behave as
$\Omega_{\rm DE} \simeq d_1/[(1+z)^6 (1+z-d_3)^2]$,
$\Omega_r \simeq (1+z)/(1+z-d_3)$, and
$\Omega_K \simeq d_2/[(1+z)(1+z-d_3)]$.

\begin{figure*}[!t]
\rotatebox{0}{\hspace{0cm}\resizebox{.45\textwidth}{!}{\includegraphics{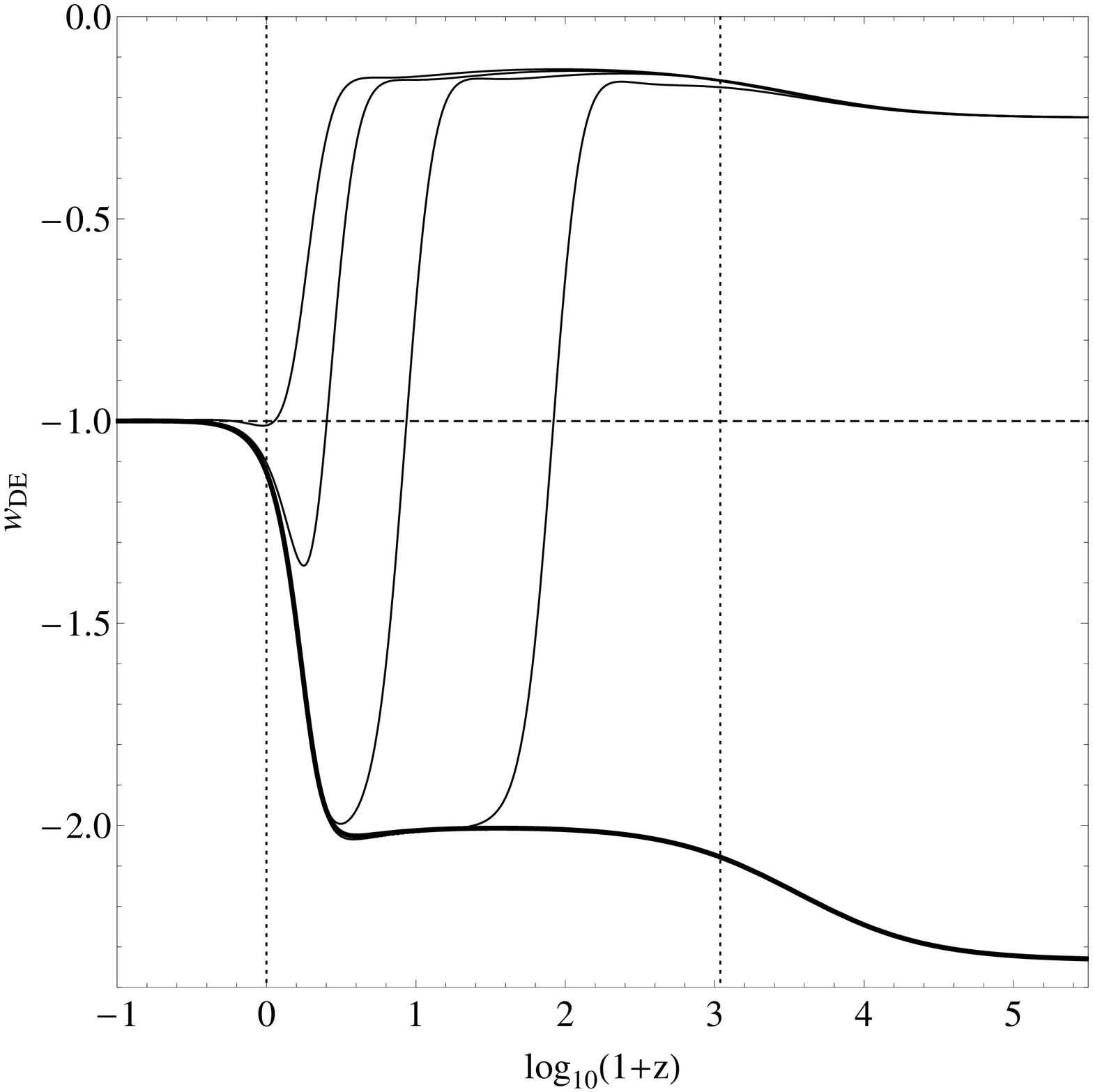}}}
\rotatebox{0}{\hspace{0cm}\resizebox{.45\textwidth}{!}{\includegraphics{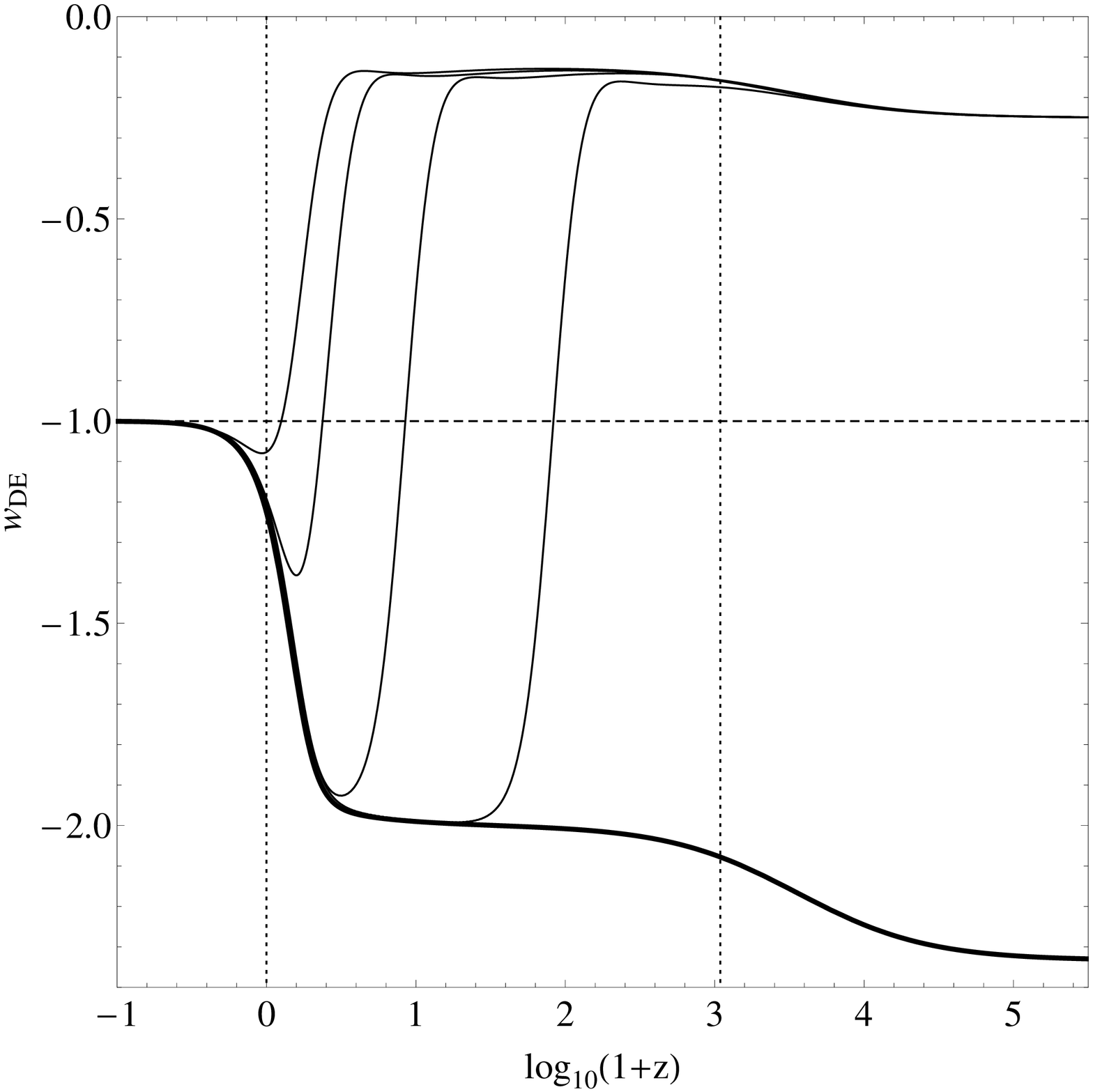}}}
\vspace{0pt}\caption{The evolution of the dark energy equation of state $w_{\rm DE}$ versus
the redshift $z$ for $\omk=-0.1$ (left) and $\omk=0.1$ (right)
with $\alpha=0.3$ and $\beta=0.14$. We choose the initial conditions
as those given in the caption of figure 3 of Ref.~\cite{DT3}.
The tracker solution ($r_1=1$) is shown as a bold line.
For given $\Omega_r (N)$, the curvature density parameter
$\Omega_K(N)$ is determined according to Eq.~(\ref{Omere}).
The two vertical lines denote the present time ($z=0$) and the epoch
at decoupling $z=z_{*}$, while the
horizontal line corresponds to the \lcdm model ($w_{\rm DE}=-1$).}
\label{fig1}
\end{figure*}

\begin{figure}[!t]
\rotatebox{0}{\hspace{0cm}\resizebox{.45\textwidth}{!}{\includegraphics{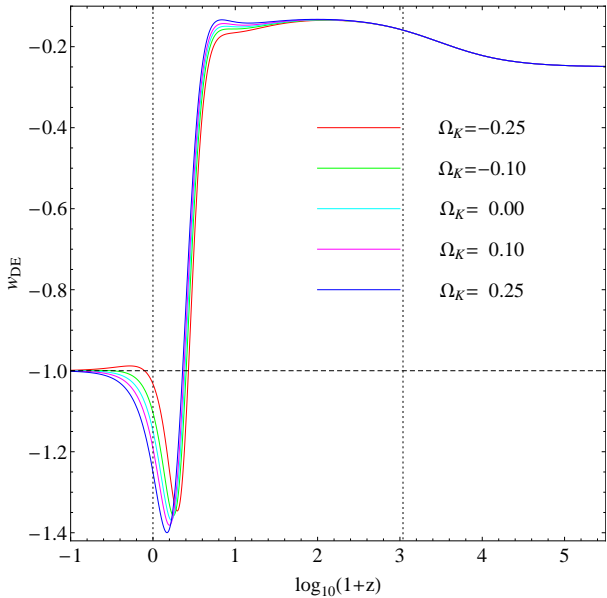}}}
\vspace{0pt}\caption{The evolution of $w_{\rm DE}$
for $\alpha=0.3$ and $\beta=0.14$ with various values of the cosmic
curvature: $\omk=(-0.25,-0.1,0,0.1,0.25)$.
The initial conditions are chosen to be
$r_1=1.5 \times 10^{-10}, r_2=2.667 \times 10^{-12}, \Omega_r=0.999992$
at $z=3.63\times10^8$ [where $N=-\ln(1+z)$].
The meaning of two vertical lines and the horizontal line
is the same as in Fig.~\ref{fig1}.}
\label{fig2}
\end{figure}

Since $\Omega_r \propto \rho_r/H^2 \propto 1/(a^4 H^2)$,
it follows that $H^2/H_0^2=(\Omega_r^{(0)}/\Omega_r)(1/a^4)$.
Using Eqs.~(\ref{eq:omsol}) and (\ref{d1d3}), the Hubble parameter
can be expressed in terms of the function of the redshift $z=1/a-1$:
\begin{widetext}
\begin{eqnarray}
  \left( \frac{H(z)}{H_0} \right)^2 & = & \frac{1}{2}
  \Omega^{(0)}_K (1 + z)^2 +
  \frac{1}{2} \Omega^{(0)}_m (1 + z)^3 + \frac{1}{2} \Omega^{(0)}_r (1 +
  z)^4 \nonumber \\
  & & +  \sqrt{1 - \Omega^{(0)}_m - \Omega^{(0)}_r - \Omega^{(0)}_K +
  \frac{(1 + z)^4}{4}  \left[ \Omega^{(0)}_K + \Omega^{(0)}_m (1 + z) +
  \Omega^{(0)}_r (1 + z)^2 \right]^2}\,.
  \label{Htracker1}
\end{eqnarray}
\end{widetext}
This analytic estimation is useful to constrain
the tracker solution from a number of observations.

On the tracker the equation of state of dark energy
$w_{\rm DE} \equiv P_{\rm DE}/\rho_{\rm DE}$ and the effective
equation of state $w_{\rm eff} \equiv -1-2\dot{H}/(3H^2)$
are given by
\begin{eqnarray}
& & w_{\rm DE}=-\frac{\Omega_r-\Omega_K+6}{3(r_2+1)}\,,\label{wdetra}
\\
& &w_{\rm eff}=\frac{\Omega_r-\Omega_K-6r_2}{3(r_2+1)}\,.
\end{eqnarray}
During the cosmological sequence of radiation ($\Omega_r \simeq 1$,
$|\Omega_K| \ll 1$, $r_2 \ll 1$), matter
($\Omega_r \ll 1$, $|\Omega_K| \ll 1$, $r_2 \ll 1$),
and dS ($\Omega_r \ll 1$, $|\Omega_K| \ll 1$, $r_2=1$)
eras, the dark energy equation of state shows peculiar evolution:
$w_{\rm DE}=-7/3 \to -2 \to -1$, whereas the effective equation
of state evolves as $w_{\rm eff}= 1/3 \to 0 \to -1$.
If the curvature density parameter $\Omega_K$ is non-negligible today,
this also gives some contribution to $w_{\rm DE}$ and $w_{\rm eff}$
during the evolution from the end of the matter era to the present epoch.
In Fig.~\ref{fig1} the evolution of the tracker solution (shown as a bold line)
for $\Omega_K^{(0)}=-0.1$ does not look very much different from that
for $\Omega_K^{(0)}=0.1$.
We caution, however, that inclusion of the cosmic curvature changes
the diameter distance as well as the luminosity distance
relative to the flat Universe.

The conditions for the avoidance of ghosts and instabilities have been
derived in Refs.~\cite{DT2,DT3} for $\Omega_K=0$.
Even in the presence of the cosmic curvature it is a good approximation
to use the results in the flat case, because $\Omega_K$ is much
smaller than 1 in most of the expansion history of the Universe.
Hence we shall use the allowed region in the $(\alpha,\beta)$ plane
shown in the figure 1 of Ref.~\cite{DT2}.

\subsection{General solutions}
\label{genesec}

There is another case in which the solutions start to evolve from
the regime $r_1 \ll 1$.
If $r_1 \ll 1$ initially, the ghosts are
absent for $\beta>0$ \cite{DT2,DT3}.
Provided that $r_1 \ll 1$ the propagation speeds of scalar and
tensor perturbations are positive during radiation and matter eras,
so that no instabilities are present.

In the regime $r_1 \ll 1$, the variables $r_1$ and $r_2$ satisfy
the following approximate equations
\begin{eqnarray}
& & r_1' \simeq \frac{9+\Omega_r-\Omega_K+21 \beta r_2}
{8+21 \beta r_2} r_1\,,\\
& & r_2' \simeq\frac{3+11\Omega_r-11\Omega_K-21\beta r_2}
{8+21\beta r_2}r_2\,.
\end{eqnarray}
As long as $\{\beta r_2, |\Omega_K| \} \ll 1$, the evolution of
$r_1$ and $r_2$ during the radiation (matter) era is given by
$r_1 \propto a^{5/4}$ and $r_2 \propto a^{7/4}$
($r_1 \propto a^{9/8}$ and $r_2 \propto a^{3/8}$).
Hence the field velocity evolves as
$\dot{\phi} \propto t^{3/8}$ during the radiation era
and $\dot{\phi} \propto t^{1/4}$ during the matter era.
Note that the evolution of $\dot{\phi}$ is slower than
that for the tracker (i.e. $\dot{\phi} \propto t$).

In Ref.~\cite{DT3} it was shown that the tracker
is stable in the direction of $r_1$ by considering a
homogeneous perturbation $\delta r_1$.
This means that once the solutions reach the tracker
the variable $r_1$ does not repel away from 1.
The epoch at which the solutions approach
the tracker regime depends on the initial values of $r_1$.

The dark energy equation of state $w_{\rm DE}$ and
the effective equation of state $w_{\rm eff}$ in the
regime $r_1 \ll 1$ are approximately given by
\begin{eqnarray}
& & w_{\rm DE} \simeq -\frac{1+\Omega_r-\Omega_K}
{8+21\beta r_2}\,, \\
& & w_{\rm eff} \simeq \frac{8\Omega_r-8\Omega_K-21\beta r_2}
{3(8+21\beta r_2)}\,.
\end{eqnarray}
Provided that $\{\beta r_2, |\Omega_K| \} \ll 1$, one has
$w_{\rm DE} \simeq -1/4, w_{\rm eff} \simeq 1/3$
during the radiation era and
$w_{\rm DE} \simeq -1/8, w_{\rm eff} \simeq 0$
during the matter era.
The evolution of $w_{\rm DE}$ in the regime $r_1 \ll 1$ is
quite different from that for the tracker solution.

In Fig.~\ref{fig1} we plot the evolution of $w_{\rm DE}$ for a number
of different initial conditions. Since $r_1 \ll 1$ initially, the solutions
start to evolve from the value $w_{\rm DE} \simeq -1/4$ in the
radiation era. For larger initial values of $r_1$
they approach the tracker earlier.
This tracking behavior occurs irrespectively of the signs of
$\Omega_K^{(0)}$. In Fig.~\ref{fig2} we find that the effect of the cosmic
curvature slightly modifies the evolution of $w_{\rm DE}$
in the low-redshift regime.

\section{Method of likelihood analysis}

In this section we show the method of our likelihood analysis
to place observational constraints on the Galileon cosmology
discussed above. The modified background cosmological evolution
in this theory affects the diameter distance to the last scattering
surface as well as the luminosity distance.
The modification from the $\Lambda$CDM model
can be tested by using the data of CMB, BAO and SN Ia.

\subsection{CMB shift parameters}
\label{cmbsec}

The positions of CMB acoustic peaks are affected by the expansion
history of the Universe from the decoupling epoch to today.
In order to quantify the shift of acoustic peaks we use the
data points $(l_a, {\cal R}, z_{*})$ of Ref.~\cite{Komatsu:2010fb} (WMAP7),
where $l_a$ and ${\cal R}$ are two CMB shift
parameters \cite{Wang,Li,Lazkoz:2007cc,Sanchez:2009ka}
and $z_{*}$ is the redshift at decoupling.
For the FLRW metric (\ref{metric}) we have \cite{Bond}
\begin{equation}
{\cal R}=\sqrt{\frac{\Omega_{m}^{(0)}}{\Omega_{K}^{(0)}}}
\sinh\left(\sqrt{\Omega_{K}^{(0)}}\int_{0}^{z_*}
\frac{\rd z}{H(z)/H_0}\right)\,.
\label{calR}
\end{equation}
Numerically it is convenient to integrate the following
equation from the redshift $z=0$ to $z=z_*$:
\begin{equation}
\frac{{\rm d}{\cal R}}{{\rm d}z}=
\frac{1}{H(z)/H_0}
\sqrt{\Omega_m^{(0)}+{\cal R}^2 \Omega_K^{(0)}}\,.
\end{equation}

The multipole $l_a$ is defined by $l_a=\pi d_a^{(c)}(z_*)/r_s (z_*)$,
where $d_a^{(c)}(z_*)={\cal R}/(H_0 \sqrt{\Omega_m^{(0)}})$ is the
comoving angular diameter distance and $r_s(z_*)
=\int_{z_*}^{\infty} {\rm d}z/[\sqrt{3(1+R_s(z))}H(z)]$
is the sound horizon at the decoupling.
Note that $R_s(z)=3\Omega_b^{(0)}/[4\Omega_{\gamma}^{(0)}(1+z)]$,
where $\Omega_b^{(0)}$ and $\Omega_{\gamma}^{(0)}$ are
the today's density parameters of baryons and photons respectively.
We neglect the contribution of dark energy and the cosmic
curvature for $z>z_*$ to estimate $r_s(z_*)$.
The dark energy density parameter can be in
fact neglected even for non-tracker solutions
with $r_1 \ll 1$,
because $\Omega_{\rm DE} \simeq 7 \beta r_2$
decreases toward the past
($r_2 \propto a^{3/8}$ during the matter era and
$r_2 \propto a^{7/4}$ during the radiation era).
It then follows that (see e.g., \cite{book})
\begin{eqnarray}
l_a &=& \left[{\rm ln}\,\left(\frac{\sqrt{R_{s}(z_*)+R_{s}(z_{\rm eq})}+
\sqrt{1+R_{s}(z_*)}}{1+\sqrt{R_{s}(z_{\rm eq})}}\right)\right]^{-1}
\nonumber \\
&& \times
\frac{3\pi}{4}\sqrt{
\frac{\Omega_{b}^{(0)}}{\Omega_{\gamma}^{(0)}}}{\cal R}\,.
\end{eqnarray}

For the redshift $z_*$ there is a fitting formula by
Hu and Sugiyama \cite{Hu:1995en}:
\begin{equation}
z_*=1048\left(1+0.00124\omega_{b}^{-0.738}\right)
\left(1+g_{1}\omega_{m}^{g_{2}}\right)\,,\label{zdec}
\end{equation}
where
$g_{1}=0.0783\omega_{b}^{-0.238}/\left(1+
39.5\omega_{b}^{0.763}\right)$,
$g_{2}=0.560/\left(1+21.1\omega_{b}^{1.81}\right)$,
$\omega_b \equiv \Omega_b^{(0)}h^2$, and
$\omega_m \equiv \Omega_m^{(0)}h^2$
($h$ correspond to the uncertainty of the Hubble
parameter $H_0$ today, i.e. $H_0=100\,h$\,km\,sec$^{-1}$\,Mpc$^{-1}$).
The redshift at the radiation-matter equality is given by
$z_{\rm eq}=\Omega_m^{(0)}/\Omega_r^{(0)}-1$.

For a flat prior, the 7-year WMAP data (WMAP7) measured
best-fit values are \cite{Komatsu:2010fb}
\begin{eqnarray}
\hspace{-.5cm}
\bm{{\bar V}}_{\rm CMB} &=& \left(\begin{array}{c}
{\bar l_a} \\
{\bar {\cal R}}\\
{z_{*}}\end{array}
\right)=
\left(\begin{array}{c}
302.09\pm 0.76 \\
1.725 \pm 0.018\\
1091.3 \pm 0.91 \end{array}
\right)\,.
\label{cmbdat}
\end{eqnarray}
The corresponding inverse covariance matrix
is \cite{Komatsu:2010fb}
\begin{eqnarray}
\bm{C}_{\rm CMB}^{-1}=\left(
\begin{array}{ccc}
2.305 & 29.698  & -1.333  \\
29.698& 6825.270& -113.180\\
-1.333& -113.180& 3.414
\end{array}
\right)\,.
\end{eqnarray}
We thus define
\begin{eqnarray}
\bm{X}_{\rm CMB} &=& \left(\begin{array}{c}
l_a-302.09\\
{\cal R} - 1.725 \\
z_{*}-1091.3\end{array}
\right)\,,
\end{eqnarray}
and construct the contribution of CMB to $\chi^2$ as
\be
\chi^2_{\rm CMB}=\bm{X}_{\rm CMB}^{T}\,
\bm{C}_{\rm CMB}^{-1}\,\bm{X}_{\rm CMB}\,.
\ee

Notice that $\chi^2_{\rm CMB}$ depends on the parameters
($\omms$, $\ob$, $h$). In the analysis of the Galileon model we use the
priors $h=0.71$ and $\ob=0.02258\,h^{-2}$ \cite{Komatsu:2010fb}.
As we shall see later on, varying $h$ does not affect the final results of our study.
The density parameter of radiation today is
\be
\Omega_r^{(0)}=\Omega_{\gamma}^{(0)}
(1+0.2271 N_{\rm eff})\,,
\label{Omer}
\ee
where $\Omega_{\gamma}^{(0)}$ is the photon density parameter
and $N_{\rm eff}$ is the relativistic degrees of freedom.
We adopt the standard values
$\Omega_{\gamma}^{(0)}=2.469 \times 10^{-5}\,h^{-2}$
and $N_{\rm eff}=3.04$ \cite{Komatsu:2010fb}.

In Ref.~\cite{Wang} Wang and Mukherjee placed observational constraints
on ${\cal R}$ and $l_a$ for several different dark energy models:
the $\Lambda$CDM model, constant $w_{\rm DE}$ models, and the models
described by the parametrization $w_{\rm DE}=w_0+w_a(1-a)$.
They showed that the resulting bounds on ${\cal R}$ and $l_a$ are
independent of the dark energy models.
While our Galileon model does not belong to the models
described above, the method using the two distance measures
${\cal R}$ and $l_a$ is expected to be trustable
as well. In fact the Galileon model is consistent
with a number of
assumptions \cite{WMAP5} that validate the analysis
using the two WMAP distance priors.

\subsection{BAO}

For the BAO we apply the maximum likelihood method \cite{Lazkoz:2007cc}
using the data points of Ref.~\cite{Percival:2009xn} (SDSS7):
\begin{eqnarray}
\bm{{\bar V}}_{\rm BAO} &=& \left(\begin{array}{c}
\frac{r_s(z_d)}{D_V(0.2)} =0.1905 \pm 0.0061  \\
\frac{r_s(z_d)}{D_V(0.35)} =0.1097 \pm 0.0036
\end{array} \right)\,,
\end{eqnarray}
where $r_s(z_d)$ is the sound horizon at the baryon
drag epoch $z_d$. For $z_d$ we use the fitting formula
by Eisenstein and Hu \cite{Eisen98}:
\begin{equation}
z_d=\frac{1291\omega_{m}^{0.251}}{1+0.659\omega_{m}^{0.828}}
\left(1+b_{1}\omega_{b}^{b_{2}}\right)\,,
\end{equation}
where $b_{1}=0.313\,\omega_{m}^{-0.419}
\left(1+0.607\,\omega_{m}^{0.674}\right)$
and  $b_{2}=0.238\,\omega_{m}^{0.223}$.
The dilation scale $D_V$ at the redshift $z$ is
\begin{equation}
D_V (z)=\left[ (1+z)^2d_A^2(z)
\frac{z}{H(z)}\right]^{1/3}\,,
\label{dv1}
\end{equation}
where $d_A(z)$ is the diameter distance defined by
\begin{equation}
d_A(z)=\frac{1}{1+z}\frac{1}{H_0 \sqrt{\Omega_K^{(0)}}}
\sinh\left(\sqrt{\Omega_{K}^{(0)}}\int_{0}^{z}
\frac{\rd \tilde{z}}{H(\tilde{z})/H_0}\right)\,.
\label{angular}
\end{equation}

This encodes the visual distortion of a spherical object due to the non-euclidianity
of the FLRW space-time, which is equivalent to the geometric mean of the
distortion along the line of sight and  two orthogonal directions.

We thus construct
\begin{eqnarray}
\bm{X}_{\rm BAO} &=& \left(\begin{array}{c}
\frac{r_s(z_d)}{{D_V(0.2)}}  - 0.1905 \\
\frac{r_s(z_d)}{{D_V(0.35)} }- 0.1097
\end{array} \right)\,,
\end{eqnarray}
and using the inverse covariance matrix \cite{Percival:2009xn}
\begin{eqnarray}
\bm{C}_{\rm BAO}^{-1} &=& \left(\begin{array}{cc}
30124 & -17227 \\
-17227 & 86977\end{array}
\right)\,.
\end{eqnarray}
The contribution of BAO to $\chi^2$ is
\begin{equation}
\chi^2_{\rm BAO}=\bm{X}_{\rm BAO}^{T} \bm{C}_{\rm BAO}^{-1}
\bm{X}_{\rm BAO}\,.
\end{equation}
\subsection{SN Ia}

The analysis of SN Ia standard candles is based on the method
described in Ref.~\cite{Lazkoz:2007cc}.
We will mainly use the Constitution SN Ia dataset of
Hicken {\it et al.} \cite{Hicken:2009dk},
which constitutes in total of 397 SN Ia.
We will also use the recently released update to the Union
set, i.e. the Union2 dataset \cite{Amanullah:2010vv}.

The SN Ia observations use light curve fitters to
provide the apparent magnitude $m(z)$ of the supernovae at peak
brightness. This is related with the luminosity distance $d_L(z)$
through $m(z)=M+5\log_{10} (d_L/10\,{\rm pc})$, where
$M$ is the absolute magnitude.
Note that the luminosity distance is given by
\begin{equation}
d_L(z)=(1+z)^2 d_A(z)\,,
\end{equation}
where $d_A(z)$ is the angular diameter distance defined in
Eq.~(\ref{angular}).
Defining the dimensionless luminosity distance as
$\bar{d}_L(z) \equiv d_L(z)/H_0^{-1}$, the theoretical
value of the apparent magnitude is
\begin{equation}
m_{\rm th}(z)={\bar M} (M, H_0) + 5 \log_{10} (\bar{d}_L (z))\,,
\label{mdl}
\end{equation}
where $\bar{M}=M-5\log_{10} h+42.38$ \cite{Lazkoz:2007cc}.

The theoretical model parameters are determined by minimizing
the quantity
\begin{equation}
\chi^2_{{\rm SN\,Ia}} (\om,p_j)= \sum_{i=1}^N
\frac{[\mu_{\rm obs}(z_i) - \mu_{\rm th}(z_i)]^2}
{\sigma_{\mu \; i}^2}\,,
\label{chi2}
\end{equation}
where $N$ is the number of the SN Ia dataset,
$p_j$ is the set of parameters to be fitted, and $\sigma_{\mu \; i}^2$
are the errors due to flux uncertainties, intrinsic dispersion of
SN Ia absolute magnitude and peculiar velocity dispersion. These
errors are assumed to be Gaussian and uncorrelated. The theoretical
distance modulus is defined as
\begin{equation}
\mu_{\rm th}(z_i)\equiv m_{\rm th}(z_i) - M
=5 \log_{10} (\bar{d}_L (z)) +\mu_0\,, \label{mth}
\end{equation}
where $\mu_0= 42.38 - 5\log_{10}h$.
The steps we have followed for the usual minimization
of (\ref{chi2}) in terms of its parameters are
described in detail in
Refs.~\cite{Nesseris:2004wj,Nesseris:2005ur,Nesseris:2006er}.

\subsection{Two information criteria}

In order to see whether the Galileon model is favored over the
$\Lambda$CDM model, we will also use the two information criteria
known as AIC (Akaike Information Criterion)
and BIC (Bayesian Information Criterion),
see Ref.~\cite{Liddle:2004nh} and references there in.
The AIC is defined as
\be
{\rm AIC} =-2\ln \mathcal{L}_{\rm max}+2k\,,
\label{aic}
\ee
where the likelihood is defined as $\mathcal{L}\propto
e^{-\chi^2/2}$, the term $-2\ln \mathcal{L}_{\rm max}$
corresponds to the minimum $\chi^2$, and $k$ is the number
of parameters of the model. The BIC is defined similarly as
\be
{\rm BIC}=-2\ln \mathcal{L}_{\rm max}+k \ln N\,,
\label{bic}
\ee
where $N$ is the number of
data points in the set under consideration.

According to these criteria a model with the smaller AIC/BIC is
considered to be the best and specifically,
for the BIC a difference of 2 is considered
as positive evidence, while 6 or more as strong evidence in favor
of the model with the smaller value. Similarly, for the AIC a
difference in the range between 0 and 2 means that the two
models have about the same support from the data as the best one,
for a difference in the range between 2 and 4 this support is
considerably less for the model with the larger AIC,
while for a difference $> 10$ the model with the larger
AIC is practically irrelevant \cite{Liddle:2004nh,Biesiada:2007um}.

\section{Observational constraints}

In this section we present the observational constraints on the
Galileon cosmology. We first consider the tracker solution and later
we study general solutions discussed in Sec.~\ref{genesec}.
In the latter case we shall explore the whole parameter space
in terms of $(\alpha,\beta)$ constrained by the conditions
for the avoidance of ghosts and instabilities
(figure 1 in Ref.~\cite{DT2}).

\subsection{The tracker}

As we showed in Eq.~(\ref{Htracker1}), the Hubble parameter for the tracker
is known as a function of the redshift $z$.
Since the value of $\omr$ is fixed from the CMB [see Eq.~(\ref{Omer})],
the tracker has only two free parameters
$\om$ and $\omk$. In what follows the priors $h=0.71$ and
$\ob=0.02258\,h^{-2}$ \cite{Komatsu:2010fb} are used
in order to simplify the analysis. We have checked that having
$h$ as a free parameter for the likelihood analysis of the CMB and BAO
does not change the results much, as the best fits are always within $1\,\%$.

We should also mention that we have not solved the full perturbation equations
for the Galileon field in order to calculate the effect on the CMB.
However, the modified growth of perturbations only affects the large-scale
CMB spectrum (the multipoles $l \lesssim 10$) through the Integrated-Sachs-Wolfe (ISW)
effect. Usually this does not provide tighter constraints than the CMB
distance measures explained in Sec.~\ref{cmbsec}.
Hence we expect that the results are not subject to change much.

\vspace{0pt}
\begin{table*}[p]
\begin{center}
\caption{Comparison of the tracker solution characterized by
(\ref{Htracker1}) to the $\Lambda$CDM model
for various data combinations.
If we use the SN Ia data alone, the model is practically
indistinguishable from the $\Lambda$CDM.
However, if we add the CMB+BAO data, the tracker solution
is disfavored relative to the $\Lambda$CDM.
In all cases the error estimation is only statistical.
In the cases where
$\omk=0$ it means that we assume the flat Universe
in the data analysis.\label{tabtrack}}
\begin{tabular}{ccccccc}
\hline
\hline\\
\hspace{6pt}\vspace{4pt}  Model (datasets)               &\hspace{6pt}$\chi^2_{\rm min}$ &\hspace{6pt}$\omk$              &\hspace{6pt} $\om$          \\
\hline \\
\hspace{6pt}\vspace{4pt}  \lcdm (CMB)                    &\hspace{6pt} $0.22$        &\hspace{6pt} $0$                &\hspace{6pt} $0.268 \pm 0.003$ \\
\hspace{6pt}\vspace{4pt}  Tracker (CMB)                  &\hspace{6pt} $16.50$       &\hspace{6pt} $0$                &\hspace{6pt} $0.316 \pm 0.004$ \\
\hspace{6pt}\vspace{4pt}  \lcdm (CMB)                    &\hspace{6pt} $0.21$        &\hspace{6pt} $0.000 \pm 0.004$  &\hspace{6pt} $0.27 \pm 0.01$ \\
\hspace{6pt}\vspace{4pt}  Tracker (CMB)                  &\hspace{6pt} $0.21$        &\hspace{6pt} $-0.016 \pm 0.003$ &\hspace{6pt} $0.27 \pm 0.01$ \\
\hline \\
\hspace{6pt}\vspace{4pt}  \lcdm (BAO)                    &\hspace{6pt} $1.58$        &\hspace{6pt} $0$                &\hspace{6pt} $0.306^{+0.047}_{-0.041}$ \\
\hspace{6pt}\vspace{4pt}  Tracker (BAO)                  &\hspace{6pt} $0.47$        &\hspace{6pt} $0$                &\hspace{6pt} $0.263^{+0.039}_{-0.034}$ \\
\hspace{6pt}\vspace{4pt}  \lcdm (BAO)                    &\hspace{6pt} $0.00$        &\hspace{6pt} $-0.37 \pm 0.23$   &\hspace{6pt} $0.22 \pm 0.06$ \\
\hspace{6pt}\vspace{4pt}  Tracker (BAO)                  &\hspace{6pt} $0.00$        &\hspace{6pt} $-0.33 \pm 0.61$   &\hspace{6pt} $0.22 \pm 0.07$ \\
\hline \\
\hspace{6pt}\vspace{4pt}  \lcdm (CMB+BAO)                &\hspace{6pt} $2.65$        &\hspace{6pt} $0$                &\hspace{6pt} $0.268 \pm 0.004$ \\
\hspace{6pt}\vspace{4pt}  Tracker (CMB+BAO)              &\hspace{6pt} $18.71$       &\hspace{6pt} $0$                &\hspace{6pt} $0.316 \pm 0.004$ \\
\hspace{6pt}\vspace{4pt}  \lcdm (CMB+BAO)                &\hspace{6pt} $2.56$        &\hspace{6pt} $0.001 \pm 0.004$  &\hspace{6pt} $0.27 \pm 0.01$ \\
\hspace{6pt}\vspace{4pt}  Tracker (CMB+BAO)              &\hspace{6pt} $0.67$        &\hspace{6pt} $-0.016 \pm 0.26$  &\hspace{6pt} $0.27 \pm 0.12$ \\
\hline \\
\hspace{6pt}\vspace{4pt}  \lcdm (Constitution)           &\hspace{6pt} $465.512$     &\hspace{6pt} $0$                &\hspace{6pt} $0.289 \pm 0.022$ \\
\hspace{6pt}\vspace{4pt}  Tracker (Constitution)         &\hspace{6pt} $465.508$     &\hspace{6pt} $0$                &\hspace{6pt} $0.392 \pm 0.024$ \\
\hspace{6pt}\vspace{4pt}  \lcdm (Constitution)           &\hspace{6pt} $465.353$     &\hspace{6pt} $-0.09 \pm 0.23$   &\hspace{6pt} $0.325 \pm 0.092$ \\
\hspace{6pt}\vspace{4pt}  Tracker (Constitution)         &\hspace{6pt} $465.423$     &\hspace{6pt} $0.076 \pm 0.258$  &\hspace{6pt} $0.358 \pm 0.118$ \\
\hline \\
\hspace{6pt}\vspace{4pt}  \lcdm (Union2)                 &\hspace{6pt} $541.011$     &\hspace{6pt} $0$                &\hspace{6pt} $0.269 \pm 0.020$ \\
\hspace{6pt}\vspace{4pt}  Tracker (Union2)               &\hspace{6pt} $541.088$     &\hspace{6pt} $0$                &\hspace{6pt} $0.369 \pm 0.022$ \\
\hspace{6pt}\vspace{4pt}  \lcdm (Union2)                 &\hspace{6pt} $540.879$     &\hspace{6pt} $-0.07 \pm 0.19$   &\hspace{6pt} $0.296 \pm 0.076$ \\
\hspace{6pt}\vspace{4pt}  Tracker (Union2)               &\hspace{6pt} $540.853$     &\hspace{6pt} $0.106 \pm 0.217$  &\hspace{6pt} $0.324 \pm 0.097$ \\
\hline \\
\hspace{6pt}\vspace{4pt}  \lcdm (Constitution+BAO)       &\hspace{6pt} $467.214$     &\hspace{6pt} $0$                &\hspace{6pt} $0.269 \pm 0.020$ \\
\hspace{6pt}\vspace{4pt}  Tracker (Constitution+BAO)     &\hspace{6pt} $472.729$     &\hspace{6pt} $0$                &\hspace{6pt} $0.367 \pm 0.022$ \\
\hspace{6pt}\vspace{4pt}  \lcdm (Constitution+BAO)       &\hspace{6pt} $466.721$     &\hspace{6pt} $-0.052 \pm 0.074$ &\hspace{6pt} $0.306 \pm 0.028$ \\
\hspace{6pt}\vspace{4pt}  Tracker (Constitution+BAO)     &\hspace{6pt} $467.066$     &\hspace{6pt} $0.172 \pm 0.067$  &\hspace{6pt} $0.311 \pm 0.029$ \\
\hline \\
\hspace{6pt}\vspace{4pt}  \lcdm (Union2+BAO)             &\hspace{6pt} $543.210$     &\hspace{6pt} $0$                &\hspace{6pt} $0.277 \pm 0.018$ \\
\hspace{6pt}\vspace{4pt}  Tracker (Union2+BAO)           &\hspace{6pt} $546.499$     &\hspace{6pt} $0$                &\hspace{6pt} $0.349 \pm 0.020$ \\
\hspace{6pt}\vspace{4pt}  \lcdm (Union2+BAO)             &\hspace{6pt} $542.013$     &\hspace{6pt} $-0.077 \pm 0.071$ &\hspace{6pt} $0.296 \pm 0.027$ \\
\hspace{6pt}\vspace{4pt}  Tracker (Union2+BAO)           &\hspace{6pt} $542.182$     &\hspace{6pt} $0.148 \pm 0.066$  &\hspace{6pt} $0.303 \pm 0.028$ \\
\hline \\
\hspace{6pt}\vspace{4pt}  \lcdm (Constitution+BAO+CMB)   &\hspace{6pt} $469.024$     &\hspace{6pt} $0$                &\hspace{6pt} $0.269 \pm 0.004$ \\
\hspace{6pt}\vspace{4pt}  Tracker (Constitution+BAO+CMB) &\hspace{6pt} $494.397$     &\hspace{6pt} $0$                &\hspace{6pt} $0.318 \pm 0.004$ \\
\hspace{6pt}\vspace{4pt}  \lcdm (Constitution+BAO+CMB)   &\hspace{6pt} $468.543$     &\hspace{6pt} $ 0.002 \pm 0.003$ &\hspace{6pt} $0.275 \pm 0.010$ \\
\hspace{6pt}\vspace{4pt}  Tracker (Constitution+BAO+CMB) &\hspace{6pt} $490.225$     &\hspace{6pt} $-0.008 \pm 0.003$ &\hspace{6pt} $0.296 \pm 0.011$ \\
\hline \\
\hspace{6pt}\vspace{4pt}  \lcdm (Union2+BAO+CMB)         &\hspace{6pt} $543.660$     &\hspace{6pt} $0$                &\hspace{6pt} $0.268 \pm 0.003$ \\
\hspace{6pt}\vspace{4pt}  Tracker (Union2+BAO+CMB)       &\hspace{6pt} $565.576$     &\hspace{6pt} $0$                &\hspace{6pt} $0.318 \pm 0.004$ \\
\hspace{6pt}\vspace{4pt}  \lcdm (Union2+BAO+CMB)         &\hspace{6pt} $543.582$     &\hspace{6pt} $ 0.001 \pm 0.003$ &\hspace{6pt} $0.270 \pm 0.010$ \\
\hspace{6pt}\vspace{4pt}  Tracker (Union2+BAO+CMB)       &\hspace{6pt} $560.183$     &\hspace{6pt} $-0.008 \pm 0.003$ &\hspace{6pt} $0.294 \pm 0.011$ \\
\hline \hline
\end{tabular}
\end{center}
\end{table*}


\begin{figure}[!t]
\rotatebox{0}{\hspace{0cm}\resizebox{.45\textwidth}{!}{\includegraphics{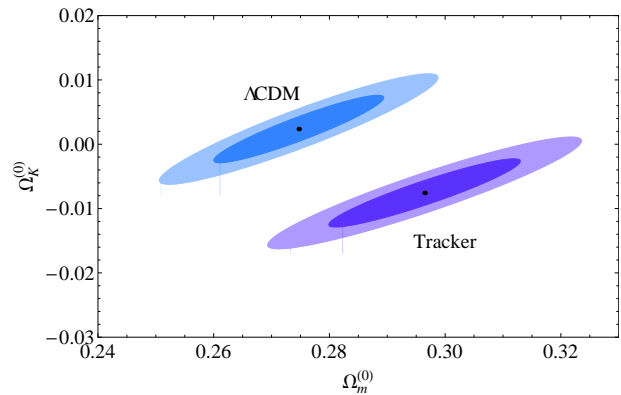}}}
\vspace{0pt}\caption{The $68.3\%\,(1\sigma)-95.4\%$\,($2\sigma$) $\chi^2$
confidence contours in the $(\om,\omk)$ plane for (i) the tracker model
and (ii) the $\Lambda$CDM model.
Both contours correspond to the combination of all three
datasets, i.e. Constitution SN Ia+BAO+CMB.
The best fit parameters are found in Table~\ref{tabtrack}.
The difference in $\chi^2$ between the two models is
$\delta \chi^2 \sim 22$, which corresponds to $\sim 4.3 \sigma$.
Hence the tracker model is severely disfavored with respect
to the \lcdm model.}
\label{fig3}
\end{figure}

In Table~\ref{tabtrack} we show a comparison of the tracker given in
Eq.~(\ref{Htracker1}) to the $\Lambda$CDM model
for various data combinations.
If we only use either of the SN Ia data (Constitution or Union2),
the $\chi^2$ for the tracker is similar to that in the \lcdm
for both the flat and non-flat cases.
Hence the two models are practically indistinguishable from
each other. The tracker solution is also consistent with the individual
observational constraint from either CMB or BAO.
The combined data analysis of CMB+BAO shows that, as long as
we take into account the cosmic curvature, the $\chi^2$ for the tracker
is even smaller than that in the $\Lambda$CDM.

In Fig.~\ref{fig3} we plot the
$\chi^2\,(=\chi^2_{\rm SN}+\chi^2_{\rm CMB}+\chi^2_{\rm BAO})$ confidence contours in the $(\om,\omk)$ plane
at the $68.3\%\,(1\sigma)-95.4\%$\,($2\sigma$) levels.
Both contours correspond to
the combination of all three datasets: Constitution SN Ia+BAO+CMB.
As we see in Table~\ref{tabtrack}, the difference
in $\chi^2$ between the two models is $\delta \chi^2 \sim 22$.
This corresponds to $\sim 4.3 \sigma$, and the tracker solution
is severely disfavored with respect to the $\Lambda$CDM.
A similar conclusion is reached from the combined data analysis
of Union2+BAO+CMB, which gives the difference of
$\delta \chi^2 \sim 16$ relative to
the $\Lambda$CDM model.

The reason why the tracker is disfavored can be explained by
inspecting Table~\ref{tabtrack} carefully.
For the flat Universe the CMB data are practically
incompatible with the tracker because of
the peculiar evolution of $w_{\rm DE}$.
The main effect on the modification of the angular diameter
distance to the last scattering surface comes from the contribution in
the low-redshift regime ($z \lesssim$ a few).
In the non-flat background the CMB data
are consistent with the tracker with the best fit values
$\om \sim 0.27$ and $\omk \sim -0.016$.
This value of $\om$ is quite smaller than the
SN Ia best fit ($\om \gtrsim 0.32$), with either
the Constitution or the Union2 datasets.

Regarding the SN Ia observations, the maximum redshift for the Constitution 
dataset is $z_{\rm max}=1.551$ and $z_{\rm max}=1.4$ for the Union2 
dataset, while the average redshifts are $z_{\rm mean}=0.4$ and 
$z_{\rm mean}=0.35$ respectively. At $z=0.4$ and $z=1.4$ the dark 
energy equations of state are given by $w_{\rm DE} \sim -1.4$ and 
$w_{\rm DE} \sim -1.9$, respectively. 
Since these values are away from $w_{\rm DE}= -1$, 
the tracker can be compatible with the SN Ia data at the expense 
of choosing the values of $\Omega_m^{(0)}$ larger than 0.3.

For the BAO datasets the best-fit value of $\om$ in the 
non-flat Universe is even smaller ($\om \sim 0.22$) than that 
constrained by the CMB shift parameters.
Thus, as each dataset favors the value of $\om$ in a different range,
the combined analysis with all datasets does not favor the
tracker solution.

\subsection{General solutions}

\begin{figure}[!t]
\rotatebox{0}{\hspace{0cm}\resizebox{.45\textwidth}{!}{\includegraphics{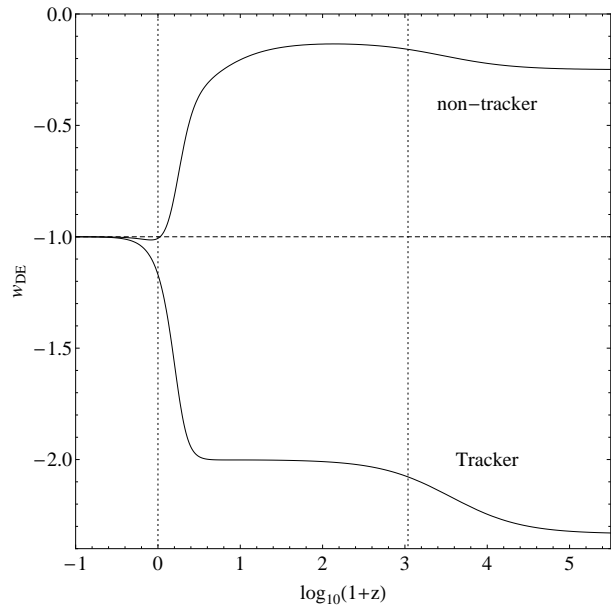}}}
\vspace{0pt}\caption{The evolution of the dark energy equation of state $w_{\rm DE}$
for the tracker solution (\ref{Htracker1}) and the non-tracker case of
Table~\ref{tabparam} [denoted by an $(*)$, i.e. $\alpha=1.862$ and $\beta=0.607$
with the initial conditions given in the text].
The meaning of two vertical lines and the horizontal line
is the same as in Fig.~\ref{fig1}.}
\label{fig4}
\end{figure}

\vspace{0pt}
\begin{table*}[!t]
\begin{center}
\caption{Comparison of general solutions of Eqs.~(\ref{eq:DRr1})-(\ref{eq:DRr4}) to
the $\Lambda$CDM model for various data combinations and values of the curvature
parameter $\omk$. For the model that is denoted by an asterisk $(*)$,
we show the evolution of $w_{\rm DE}$ in Fig.~\ref{fig4}
(labeled as ``non-tracker'').
\label{tabparam}}
\begin{tabular}{ccccccc}
\hline
\hline\\
\hspace{1pt}\vspace{6pt}   Model/Data
&\hspace{6pt}$\chi^2_{\rm min}$ & \hspace{6pt}Best fit parameters & AIC &$\Delta$AIC&AIC&$\Delta$BIC  \\
\hline \\
\hspace{1pt}\vspace{6pt}(Constitution+BAO+CMB)             & & & & & & \\
\hline
\hspace{1pt}\vspace{6pt}  $ \Lambda$CDM $(\omk=0)$         &\hspace{6pt} $469.024$ &\hspace{6pt} $\om=0.269\pm0.004$  &471.024&0&475.020&0\\
\hspace{1pt}\vspace{6pt}  $\Lambda$CDM  $(\omk\neq0)$      &\hspace{6pt} $468.543$ &\hspace{6pt} ${}_{\om=0.267\pm0.004}^{\omk=0.002\pm0.003}$  &472.543&1.519&480.536&5.515\\
\hspace{1pt}\vspace{6pt}  $(\alpha, \beta)$, $(\omk=0)$     &\hspace{6pt} $470.912$ &\hspace{6pt} ${}_{\beta=0.422 \pm 0.022}^{\alpha=1.411 \pm 0.056}$ &474.912&3.888&482.905&7.884\\
\hspace{1pt}\vspace{6pt}  $(\alpha, \beta)$, $(\omk=-0.01)$ &\hspace{6pt} $537.543$ &\hspace{6pt} ${}_{\beta=0.477 \pm 0.027}^{\alpha=0.956 \pm 0.079}$ &541.543&70.519&549.536&74.515\\
\hspace{1pt}\vspace{6pt}  (*) $(\alpha, \beta)$, $(\omk=0.01)$  &\hspace{6pt} $468.545$ &\hspace{6pt} ${}_{\beta=0.607 \pm 0.024}^{\alpha=1.862 \pm 0.058}$ &472.545&1.521&480.538&5.517\\
\hspace{1pt}\vspace{6pt} $(\alpha, \beta, \om, \omk)$  &\hspace{6pt} $468.311$ &\hspace{6pt} ${}_{\beta=0.425 \pm 0.064}^{\alpha=1.401 \pm 0.159}$ &476.311&5.287&492.297&17.276\\
\hspace{1pt}\vspace{6pt}  & &\hspace{6pt} ${}_{\omk=-0.003 \pm 0.005}^{\om=0.287 \pm 0.014}$ & & & & \\
\hline\\
\hspace{1pt}\vspace{6pt}(Union2+BAO+CMB)
 & & & & & & \\
\hline
\hspace{1pt}\vspace{6pt}  $\Lambda$CDM $(\omk=0)$&\hspace{6pt} $543.660$ &\hspace{6pt} $\om=0.268 \pm 0.003$  &545.660&0&549.992&0\\
\hspace{1pt}\vspace{6pt}  $\Lambda$CDM $(\omk\neq0)$ &\hspace{6pt} $543.582$ &\hspace{6pt} ${}_{\om=0.270\pm0.010}^{\omk=0.001\pm0.003}$  &547.582&1.922&556.245&6.254\\
\hspace{1pt}\vspace{6pt}  $(\alpha,\beta)$, $(\omk=0)$      &\hspace{6pt} $543.895$ &\hspace{6pt} ${}_{\beta=0.419 \pm 0.023}^{\alpha=1.404 \pm 0.057}$ &547.895&2.235&556.558&6.567\\
\hspace{1pt}\vspace{6pt}  $(\alpha,\beta)$, $(\omk=-0.01)$  &\hspace{6pt} $610.919$ &\hspace{6pt} ${}_{\beta=0.477 \pm 0.027}^{\alpha=0.956 \pm 0.089}$ &614.919&69.259&623.582&73.591\\
\hspace{1pt}\vspace{6pt}  $(\alpha,\beta)$, $(\omk=0.01)$   &\hspace{6pt} $543.646$ &\hspace{6pt} ${}_{\beta=0.603 \pm 0.024}^{\alpha=1.855 \pm 0.058}$ &547.646&1.986&556.309&6.318\\
\hspace{1pt}\vspace{6pt} $(\alpha, \beta, \om, \omk)$  &\hspace{6pt} $542.290$ &\hspace{6pt} ${}_{\beta=0.423 \pm 0.313}^{\alpha=1.401 \pm 0.779}$ &550.290&4.630&567.616&17.625\\
\hspace{1pt}\vspace{6pt}  & &\hspace{6pt} ${}_{\omk=-0.002 \pm 0.019}^{\om=0.282 \pm 0.016}$ & & & & \\
\hline \hline
\end{tabular}
\end{center}
\end{table*}

Next, we proceed to observational constraints on the general solutions
to Eqs.~(\ref{eq:DRr1})-(\ref{eq:DRr4}).
Unlike the tracker solution, this case depends on the parameters
$\alpha$ and $\beta$ as well as the initial conditions of
$r_1$, $r_2$, and $\Omega_r$.
Note that the initial condition of $\Omega_K$ is fixed
by using Eq.~(\ref{Omere}).
{}From Eq.~(\ref{HN}) the Hubble parameter $H(N)$ is
integrated to give
\be
\frac{H(N)}{H_0}=\left(\frac{r_1(0)}{r_1(N)}\right)^{5/4}
\left(\frac{r_2(0)}{r_2(N)}\right)^{1/4}\,,
\label{HN1}
\ee
which can be used to compare the model with the data.

We place observational bound on ($\alpha$, $\beta$)
constrained by the conditions for the avoidance of ghosts and instabilities.
We also restrict $\beta>0$ to avoid the appearance of ghosts
at the early cosmological epoch \cite{DT2,DT3}.
Since the tracker is disfavored by the data, the cosmological evolution
in which the solutions approach the tracker at late times (i.e.
smaller initial values of $r_1$) is in general favored.
For given $\alpha$ and $\beta$, we search for the viable initial
conditions of $r_1$, $r_2$, and $\Omega_r$ that can be consistent
with the data. We then carry out the likelihood analysis by varying
$\alpha$ and $\beta$ to find the viable parameter space.

In Table~\ref{tabparam} we show the best fit and the best
fit parameters along their 1$\sigma$ errors and the AIC/BIC
for several fixed values of $\omk$\,($=0, -0.01, 0.01$).
According to the AIC statistics, the $\Lambda$CDM and
the non-tracker solution with the
positive curvature parameter $\omk=0.01$
(with the best-fit values $\alpha=1.862$ and $\beta=0.607$)
have more or less the same support. 
For this set of best-fit parameters we can always find a set of initial conditions 
for $r_1$ and $r_2$ that lead to the late-time tracker allowed by observations. 
For example, for the parameters $\alpha=1.862$ and $\beta=0.607$, 
the initial conditions giving rise to the late-time tracker allowed by observations 
are $r_1=1.50 \times 10^{-10} \pm 3.88 \times 10^{-12}$, 
$r_2=2.67 \times 10^{-12} \pm 2.17 \times 10^{-13}$ (95 \% CL), for 
$\Omega_r=0.999992$ at the redshift $z=3.63 \times 10^8$. 
These conditions can be translated into the initial condition 
for $\dot{\phi}$ by using Eq.~(\ref{r1r2def}). 
In the case of $\alpha=1.862$ and $\beta=0.607$
it follows that $4.38 \times 10^{-6}<\dot{\phi}/\dot{\phi}_{\rm dS} <4.57 \times 10^{-6}$ at $z=3.63 \times 10^8$, thus requiring a certain amount of fine tuning. 

For $\omk=-0.01$ the data support the $\Lambda$CDM model more, 
as the fit is particularly bad. On the other hand, according to the BIC statistics,
there seems to be strong evidence in favor of the flat \lcdm
over all non-tracker cases as $\Delta \textrm{BIC}\gtrsim 6$.

In Fig.~\ref{fig4} we illustrate the evolution of $w_{\rm DE}$ for the
solution denoted by an asterisk in Table~\ref{tabparam},
together with the tracker solution.
The former corresponds to the model parameters
$\alpha=1.862$ and $\beta=0.607$ with the initial conditions
$r_1=1.5 \times 10^{-10}, r_2=2.667 \times 10^{-12}$,
and $\Omega_r=0.999992$
at $z=3.63 \times 10^8$.
In this case the solution does not reach the tracker by today
and hence it does not exhibit strong phantom behavior.
We find that the early tracking behavior is in fact disfavored
by the combined data analysis.

The results in Table~\ref{tabparam} for fixed $\omk$
imply that the generic solutions fare
quite well for the open Universe (i.e. positive $\omk$), whereas
they are not in good agreement with the data
for the closed Universe with $\Omega_K^{(0)} \lesssim -0.01$.
However, if we perform a general fit where
all 4 parameters $(\alpha, \beta, \om, \omk)$ are allowed
to vary, then a slightly negative value of $\omk$ is favored, e.g.,
$\omk=-0.003 \pm 0.005$ for the Constitution set.
The reason why the general solutions are not particularly favored
in the AIC and BIC tests lies in the fact that the number of
parameters is larger than those in the flat $\Lambda$CDM.
Therefore, we conclude that the general solutions exhibit rich
phenomenology that can in general be in good agreement with
the combined observational constraints, while parts of the
parameter space being in mild tension with the observations.

\section{Conclusions}

In this paper we have placed observational constraints
on the covariant Galileon cosmology by using the most recent
data of SN Ia (Constitution and Union2 sets), CMB (WMAP7),
and BAO (SDSS7). In this theory there is an interesting
tracker solution that finally approaches a de Sitter solution
responsible for dark energy.
By including the cosmic curvature $\Omega_K^{(0)}$,
we derived the analytic formula (\ref{Htracker1}) about the
evolution of the Hubble parameter for the tracker.
This formula is convenient because we only need to vary the two
density parameters $\Omega_m^{(0)}$ and $\Omega_K^{(0)}$
for the likelihood analysis as in the $\Lambda$CDM model
(with the radiation density parameter $\Omega_r^{(0)}$
fixed by the CMB).

If we use either of the SN Ia data (Constitution or Union2) alone,
the $\chi^2$ for the tracker is similar to that in the $\Lambda$CDM model.
We also found that, as long as the cosmic curvature is taken into account,
the tracker solution is compatible with the individual
observational bound constrained from either CMB or BAO.
However, the combined data analysis of Constitution+BAO+CMB shows
that the difference of $\chi^2$ between the tracker and the $\Lambda$CDM
is $\delta \chi^2 \sim 22$ (or $\sim 4.3 \sigma$).
Hence the tracker is severely disfavored with respect to the $\Lambda$CDM.
{}From the combined data analysis of Union2+BAO+CMB, we reached
a similar conclusion: the difference of
$\delta \chi^2 \sim 16$ relative to the $\Lambda$CDM.
The reason for this incompatibility is that the SN Ia data favor the large values
of $\Omega_m^{(0)}$~($\gtrsim 0.32$), whereas the CMB and BAO
data constrain smaller values of $\Omega_m^{(0)} (\lesssim 0.27)$.

We also studied the general solutions to Eqs.~(\ref{eq:DRr1})-(\ref{eq:DRr4})
for the parameters $\alpha$ and $\beta$ constrained by the conditions
for the avoidance of ghosts and instabilities.
In general the solutions that approach the tracker
only at late times
are favored from the combined data analysis.
By choosing several fixed values of $\omk$, we found that
the generic solutions can be consistent with the data for the
open Universe ($\Omega_K^{(0)}>0$).
For example, the general solutions with $\omk=0.01$ and the model
parameters $(\alpha, \beta)=(1.862, 0.607)$ give the similar value
of $\chi^2$ to that in the the $\Lambda$CDM.
In this case the AIC statistics also have the same support for
the two models.
For the models with largely negative $\omk$ such as
$\omk \lesssim-0.01$ the data favor considerably
more the \lcdm model, as the fit is particularly bad.

The BIC statistics show that  the general solutions, with all 4 parameters
$(\alpha, \beta, \om, \omk)$ are varied, are not particularly favored
over the $\Lambda$CDM model (because $\Delta \textrm{BIC}\gtrsim 6$).
This mainly comes from the statistical property that the number of
model parameters is larger than those in the flat $\Lambda$CDM.
In fact the general solutions with a non-zero curvature can be well consistent
with the combined data analysis.

It will be of interest to study the evolution of matter density perturbations
to confront the Galileon cosmology with the observations of large scale structure (LSS).
In particular, the presence of the field derivative couplings with $R$ and $G_{\nu \rho}$
in the expression of ${\cal L}_4$ and ${\cal L}_5$
will change the effective gravitational
coupling \cite{DKT}.
This should modify the growth rate of matter perturbations as well as
the ISW effect on large-scale CMB anisotropies.
Compared to the distance measures ${\cal R}$ and $l_a$ we discussed in this paper,
the ISW effect is in general not so powerful to constrain dark energy models.
However, in the case where the modification of
the effective gravitational coupling is significant,
the ISW effect may provide a strong constraint.
In addition, the LSS-ISW anti-correlation found in a similar Galileon-like
model \cite{Kobayashi2} can be
a useful tool to constrain the parameter space $(\alpha,\beta)$ further.
We will leave this issue for a future work.

\section*{ACKNOWLEDGEMENTS}
The authors are grateful to Eiichiro Komatsu, Kazuya Koyama,
Shuntaro Mizuno, Pia Mukherjee, and David Wands
for useful discussions.
S.\,N.\ is supported by the Niels Bohr International Academy,
the Danish Research Council under FNU Grant No. 272-08-0285
and the DISCOVERY center.
A.\,D.\,F.\ thanks Burin Gumjudpai for warm hospitality
during his stay in Naresuan University.
S.\,T.\ is also grateful to Savvas Nesseris, Kazuya Koyama,
Burin Gumjudpai, and Jungjai Lee for warm hospitalities
during his stays in the Niels Bohr
Institute, the University of Portsmouth, Naresuan University,
and Daejeon.
The work of A.\,D.\,F.\ and S.\,T.\ was supported by the
Grant-in-Aid for Scientific Research Fund of the
JSPS Nos.~09314 and 30318802.
S.\,T.\ also thanks financial support for the Grant-in-Aid for
Scientific Research on Innovative Areas (No.~21111006).


\end{document}